\documentclass[final,3p,times,twocolumn,showpacs,superscriptaddress]{elsarticle}
\usepackage{epsfig}
\usepackage{amssymb}
\usepackage{amsmath}
\usepackage{textcomp}
\usepackage{lineno}
\usepackage{hyperref}
\biboptions{sort&compress}

\usepackage{bm}% bold math
\usepackage{amsmath}
\usepackage{xcolor}
\usepackage[utf8]{inputenc}

\newcommand{\ee}{ {\mathrm e} }%

\numberwithin{equation}{section}
\journal{Physics of the Dark Universe}
\newcommand{\eq}{\begin{equation}}
\newcommand{\eqe}{\end{equation}}

\begin{document}

\begin{frontmatter}

\title{ %1) A dark fluid hydrodynamical model with gravitation \\ 
%2) Hydrodynamics of a self-gravitating dark-matter universe \\
Scaling Hydrodynamical Evolution of a Gravitating Dark-fluid Universe}


\author{Imre Ferenc Barna}
\ead{barna.imre@wigner.hu}
\author{Mih\'aly Andr\'as Pocsai}
\ead{mihaly.pocsai@wigner.hu}
\author{Gergely G\'abor Barnaf\"oldi}
\ead{barnafoldi.gergely@wigner.hu}

\address{Wigner Research Centre for Physics\\
29-33 Konkoly-Thege Mikl\'os Str., H-1121, Budapest, Hungary}

\begin{abstract}
We present a dark fluid model which contains the general linear equation of state including the gravitation term. The obtained spherical symmetric Euler equation and the continuity equation was investigated with the Sedov-type time-dependent self-similar {\it ansatz} which is capable to describe physically relevant diffusive and dispersive solutions. %The role of the parameter in the equation of state is investigated.  
As results the space and time dependent fluid density and radial velocity fields are presented and analyzed. Additionally, the role of the initial velocity on the kinetic and total energy densities of the fluid is discussed.   
\end{abstract}


%\item Research highlight 

\end{frontmatter}
%%\linenumbers
\section{Introduction}
\label{sec:intro}

The general being and detailed properties of the dark matter is one of the most stimulating question in astro-particle physics and cosmology. However, the first hint for the existence of such materials is by F.~Zwicky from 1933~\cite{zwicky}, the experimental evidence came more than 40 years later in the 70s from different groups~\cite{exp1,exp2,exp3}. The physics of dark matter is an interesting and active research area, which is intensively studied~\cite{bertone,sanders,debasich,dark-imre}. 
%One and a half decade ago one of us presented an atomphysical calculation which might help to decide if the dark mater is a hypothetical  particle which might interact with atomic nuclei via a Dirac delta interaction potential \cite{}.

Einstein equations can be solved with fluid energy-momentum tensor and some of them are possible candidate for the description of dark (fluid) matter. We investigate one of the simplest polytropic equation of state (EoS), the linear one, which can be inserted into the spherical Euler equation. So far, there is no general mathematical technique to ascertain all solutions and properties of non-linear partial differential equations (PDE) or systems, however there are some methods which give us a glimpse into some kind of solutions. 

One of the main direction of this research is to find scaling solutions of the gravitational fields, which can be good candidates to describe the evolution of the Universe or collapse of compact astrophysical objects even in multi-dimensional space-time~\cite{Brandt:2001,Brandt:2003,HeydariFard:2009,Worrakit:2015}. In these self-similar scaling models the time evolution of the scaling is usually restricted by the metric. Another proposed analytic solution method is the self-similar {\it ansatz} by L.~Sedov~\cite{sedov} which usually provide to evaluate physically reasonable solutions with dispersive features and with asymptotic power-law decays. This {\it ansatz} has been already applied successfully in some other hydrodynamical solutions~\cite{imre3,imre4,imre5}. 

There are some studies available which investigate the stability (or other properties of) various relativistic or non-relativistic 	gravitating fluids~\cite{taben,christ,duco,ahmad,jordano}. In their monograph Deruelle and Uzan~\cite{nath} analyse gravitating fluids and present some solutions as well. Due to our knowledge there is no time-dependent self-similar solutions known and discussed for any kind of dark fluid hydrodynamical model. Our motivation here is to investigate this dispersive solution in relation with the evolution of a gravitating system. Moreover, our results are compatible with the ideas of dark matter powered evolution of the early Universe~\cite{Dimitar,Peacock,Hinshaw} or it can be also applied to celestial dark matter objects~\cite{Freese}.

In this paper a typical dark matter fluid EoS is applied, and we study the behavior and the physical relevance of the self-similar numeric hydrodynamic solutions for specific cases in the Newtonian approximation~\cite{Csabai}. Finally, we provide the velocity, density and kinetic energy density profiles for the space-time evolution. 

%%%%%%%%%%%%%%%%%%%%%%%%%%%%%%%%%%%%%%%%%%%%%%%%%%%%%%%%%%%%%%%%%%%%%%%%%%
\section{The model} 
\label{sec:model}

Let's assume, one-dimensional, spherical symmetric system, which is described by a compressible continuity and Euler coupled PDE respectively,   
\begin{eqnarray}\label{eq:euler}
\rho_t + u_r\rho + u\rho_r + \frac{2u\rho}{r} =0, \nonumber \\
u_t + uu_r = -\frac{1}{\rho} p_r -  \left(\frac{\rho}{r}\right)_r  .  
\label{pde}
\end{eqnarray}
Here the subscripts $t$ and $r$ mean the corresponding time- and spatial partial derivatives.  
Additionally, $\rho=\rho(r,t)$, $u=u(r,t)$, and $p=p(r,t)$ mean the density, the radial velocity component and the pressure field distributions, respectively. Note, the speed of the light and the gravitational coupling constant were set to be unity.

The second term in the Euler equation~\eqref{eq:euler} on the right hand side is the gravitating term:
the radial component of the gradient of the Newtonian potential. 

We apply the general linear EoS, which has other names in different scientific communities like barotropic or "stiff matter" EoS	
\begin{equation}\label{eq:cg}
p = w \rho^n  \ \ \ \ \textrm{for} \ \ \ \ n = 1.
\end{equation} 
There are numerous EoS available for fluids or in astrophysics, for more see  
Emden \cite{emden}. He was the first who investigated polytropic EoS inside stars at the 
beginning of the 20\textsuperscript{th} century. 
The physics of various polytropic EoS in astrophysics can be found in the monograph~\cite{horedt}. 
%%%%%%%%%%%%%%%%%%%%%%%%%%%%%%%%%%%%%%%%%%%%%%%%%%%%%%%%%%%%%%%%%%%%%%%
%The case of  $n = -1$ is the so called the Chaplygin gas and if $ -1 \le n < 0 $ is the generalized Chaplygin gas. So far it is not clear, whether they can be used as models to describe the totality (or the majority) of dark matter in the universe. 
%On one hand, both models are ruled out as candidates for the dark matter fluid~\cite{chap,gen_chap}, since the pressure gradients of these fluids is nonzero and if they are used as dark matter they fail to reproduce standard cosmological observations such as the CMB anisotropies or the baryonic acoustic oscillations~\cite{sandvik}. On the other hand the Chaplygin gas is the only fluid which, upto now, admits a supersymmetric generalization~\cite{debasich,dark-imre}. We found that to obtain stabilization it is necessary to add matter on the branes which again obeys the equation of state.
%
%Indeed Eq.~\eqref{eq:cg} has an amusing connection with string theory and it can be obtained from the Nambu\,--\,Goto action for $d$-branes moving in a $(d+2)$-dimensional spacetime in the light-cone parametrization~\cite{sanders}.
%
%
%%%%%%%%%%%%%%%%%%%%%%%%%%%%%%%%%%%%%%%%%%%%%%%%%%%%%%%%%
The adiabatic speed of sound we now automatically get the constant of 
\begin{equation}
\frac{d p}{d \rho} = c_s^2 = w   
\end{equation}
which is a necessary physical condition. Different numerical values of the EoS strength $w$ may lead to different dark matter scenarios for negative values as it was outlined by Perkov~\cite{perkov}. 
For the later understanding of our calculations we give some examples according to~\cite{hogan, vikman}:\\ 
\begin{enumerate} 
\item[(i)] {
$w = 0$ means the EoS for ordinary non-relativistic 'matter' (e.g. cold dust)}; 
\item[(ii)]{$w = 1/3$ means ultra-relativistic 'radiation' (including neutrinos) and in the very early universe other particles that later become non-relativistic}; 
\item[(iii)]{$w = -1$ is the simplest case and describes expanding universe, hypothetical phantom energy $w < - 1$ would cause Big Rip};
\item[(iv)]{$w \ne -1$ means quintessence as hypothetical fluid};
\item[(v)]{$w = -1/3$ is responsible for the flatness of the Big Bang};
\item[(vi)]{A scalar field $\phi$ can be viewed as a sort of perfect fluid with EoS of
\begin{equation} 
w = \frac{ \frac{1}{2}\phi^2_t - U({\phi}) }{\frac{1}{2}\phi^2_t + U({\phi})  }
\end{equation}
where $ \phi_t $ is the time derivative of $\phi$ and $U(\phi)$ is the potential energy. A free, $U(\phi)=0$ scalar 
field has $w = -1$ an the one with vanishing kinetic energy is equivalent to $w = 1$. Any EoS in between but not 
crossing the $w\neq-1$ barrier is known as the Phantom Divide Line (PDL)~\cite{vikman} is achievable, 
which make scalar fields useful models for any phenomena in cosmology.}   
\end{enumerate}

The Universe has gone through three distinct eras  characterized by the $z$ red shift parameter values: radiation-dominated at $z \ge 3000 $; matter-dominated if $3000 \ge z \ge 0.5$; and dark-energy-dominated when $z \le 0.5$. The evolution of the scale factor is controlled by the dominant energy form based on the Friedmann\,--\,Lema\^itre\,--\,Robertson\,--\,Walker equations: $ a(t) \propto  t^{2/3}(1 + w) $ (for constant $w$). During the radiation-dominated era, $ a(t) \propto t^{1/2}$, while in the matter-dominated era, $a(t) \propto t^{2/3}$. Finally for the dark energy-dominated era one can assume $w = -1$, which is asymptotically $ a(t) \propto \ee^{Ht}$~\cite{frieman}. In our solution we will consider most of the above cases, which lead to the same solution family, but our main focus will be on the case with $w = -1$.

%%%%%%%%%%%%%%%%%%%%%%%%%%%%%%%%%%%%%%%%%%%%%%%%%%%%%%%%%%%%%%%%%%%%%%%%%%%%%%%%%%%%
\section{Analytic solution with the Sedov-ansatz} 
\label{sec:solution}

We want to find and analyze disperse analytic solutions with the application of the well-known self-similar {\it ansatz}~\cite{sedov,barenb,zeld} of 
\eq 
V(x,t)=t^{-\alpha}f\left(\frac{x}{t^\beta}\right):=t^{-\alpha}f(\omega),  
\label{self}
\eqe 
where $V(x,t)$ can be an arbitrary variable of a PDE where $t$ means time and $x$ means spatial 
dependence. The function $f(\omega)$ is called the shape function. 
The similarity exponents $\alpha$ and $\beta$ are of primary physical importance since $\alpha$  
represents the rate of decay of the magnitude $V(x,t)$, while $\beta$  is the rate of spread  
(or contraction if $\beta<0$) of the space distribution as time goes on. 
Solutions with integer exponents are called self-similar solutions of the first kind
(and sometimes can be obtained from dimensional analysis of the problem). 
The {\it ansatz} can be generalized considering real and continuous functions $a(t)$ 
and $b(t)$ instead of $t^{\alpha} $ and $t^{\beta}$.   
%%%%%%%%%%%%%%%%%%%%%%%%%%%%%%%%%%%%%%%%%%%%%%%%%%%%
\begin{figure}[!h]
\begin{center}
\scalebox{0.38}{
\rotatebox{0}{\includegraphics{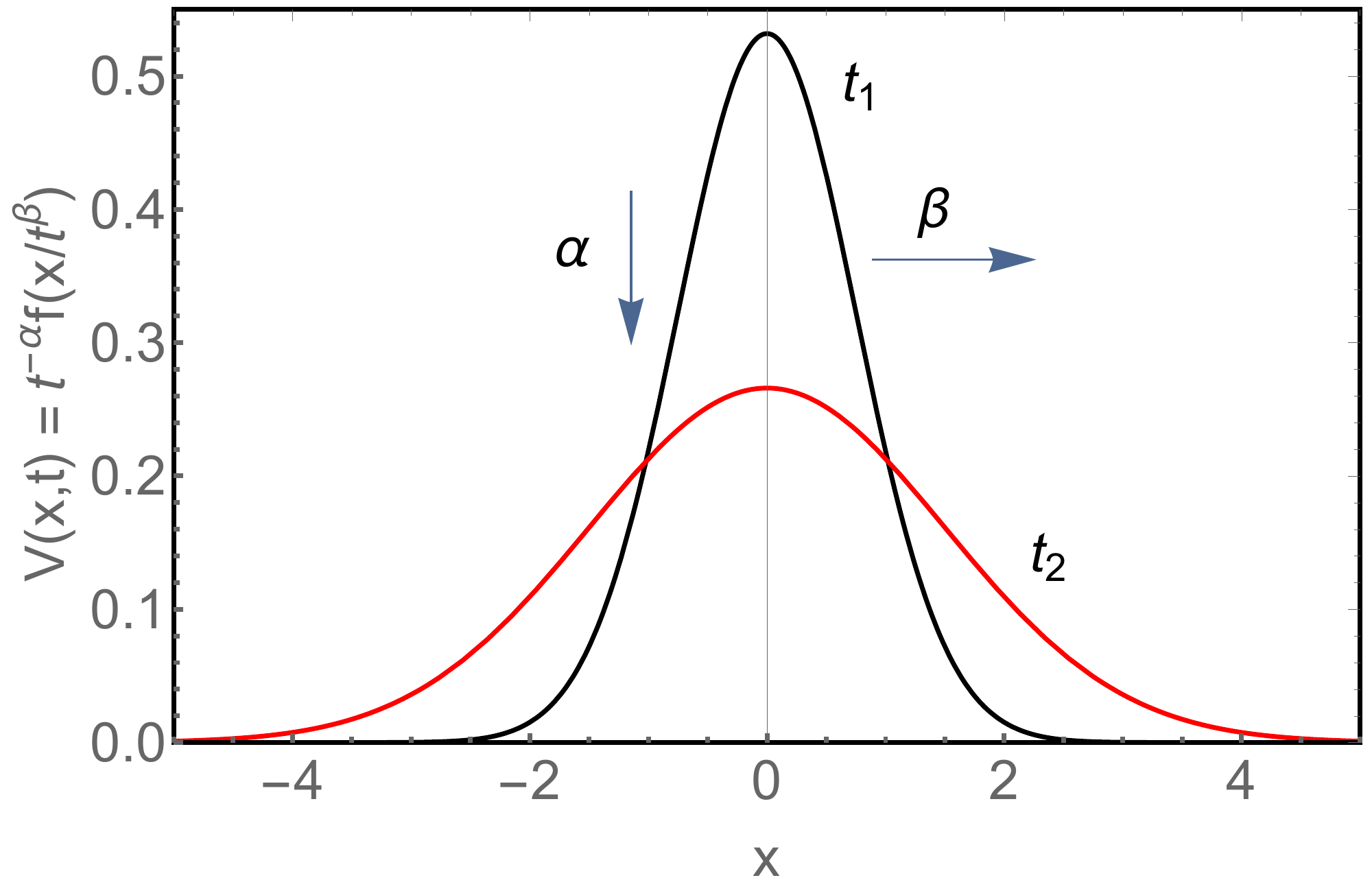}}}
\vspace*{-0.5cm}
\end{center}
\caption{A self-similar solution of Eq. (\ref{self}) for $t_1<t_2$.
The presented curves are Gaussians for regular heat conduction.}	
\label{egyes}       % Give a unique label
\end{figure}
%%%%%%%%%%%%%%%%%%%%%%%%%%%%%%%%%%%%%%%%%%%%%%%%%%

The most powerful result of this {\it ansatz} is the fundamental- or 
Gaussian-solution of the Fourier heat conduction equation (or for Fick's
diffusion equation) with $\alpha =\beta = 1/2$, which is clearly presented 
on Figure~\ref{egyes}. 
This transformation is based on the assumption that a self-similar solution
exists, i.e., every physical parameter preserves its
shape during the expansion. Self-similar solutions usually
describe the asymptotic behavior of an unbounded or a far-field
problem; the time $t$ and the space coordinate $x$ appear
only in the combination of  $f(x/t^{\beta})$.  It means that the existence
of self-similar variables implies the lack of characteristic
length and time scales. These solutions are usually not unique and
do not take into account the initial stage of the physical expansion process.
These kind of solutions describe the intermediate asymptotic of a problem: they hold when the precise initial 
conditions are no longer important, but before the system has reached its final steady state. 
For some systems it can be shown that the self-similar solution fulfills the source type 
(Dirac-delta) initial condition, but not in our next case. 
They are much simpler than the full solutions and so easier to understand and study in 
different regions of parameter space. A final reason for studying 
them is that they are solutions of a system of ordinary differential equations and hence do not suffer from the extra inherent numerical problems of the full PDEs. In some cases self-similar solutions helps to understand diffusion-like properties 
or the existence of compact supports of the solution. 

In the last decade we successfully applied this {\it ansatz} for numerous physical systems like heat 
conduction~\cite{imre1}, non-linear Maxwell equation~\cite{imre2} especially multidimensional 
Euler and Navier\,--\,Stokes equations~\cite{imre3,imre4} or the Madelung\,--\,Schr\"odinger 
quantum fluid equations~\cite{imre5} ending up with Refs.~\cite{imrebook,imrebook2}. 
 
For our system we apply the following notations for the two shape functions in natural units
\eq
u =  t^{-\alpha}f\left(\frac{r}{t^{\beta}} \right)   \ \ \ \ \textrm{and} \ \ \ \
\rho = t^{-\gamma}g\left(\frac{r}{t^{\beta}} \right), \>\> 
\label{ans}
\eqe
where the new variable is $\omega= r/t^{\beta}$. 

Calculating time and spatial derivatives of the eqs.~\eqref{ans} and substituting to eqs.~\eqref{pde} and using the relation, 
$ r = \frac{r}{t^{\beta}} t^{\beta}= \omega t^{\beta}$, and
finally we get the next 
non-linear ordinary differential equation (ODE) system
\begin{eqnarray}  
 g -  \omega g' + f'g + fg'+ \frac{2gf}{\omega} = 0, \nonumber \\
- \omega f' + ff' = - w\frac{g'}{g} -  \frac{g'}{\omega} + \frac{g}{\omega^2} , 
\label{ODE} 
\end{eqnarray}
where prime means derivation with respect to $\omega$. 

For the initially free three self-similar exponents $\alpha$, $\beta$, and $\gamma$ we obtained the following numerical values:   
$ \alpha = 0$,  $\beta = 1 $, and $\gamma = -1 $.
This means that dynamical variables, velocity and density, have spreading 
property as time goes on  ($\beta = 1$). 
Our physical intuition says that spreadings are somehow similar to expansion which is a 
basic property of the Universe. 
The two "decay" exponents ($\alpha$ and $\gamma$) are however not positive, which means that 
the magnitude of the velocity, $u$ remains the same even for large times. Density, $\rho$ is even more peculiar, the magnitude linearly enhances in time.  
We note the usual decaying and dispersive solutions with zero asymptotic values.  

The analysis of the relations among the self-similar exponents can end up with three different scenarios:
\begin{itemize}
\item{
The linear algebraic equation system among the exponents are overdetermined, 
which automatically means contradiction. Therefore the system has inherently no physically self-similar 
power-law decaying or exploding solutions. Such systems are rare but some damped wave equations e.g. telegraph equations are so.} 
\item{ All exponents have well-defined numerical values, the analysis of the solutions is 
straightforward, the remaining coupled non-linear ODE system can be analyzed, in some lucky cases 
even it can be decoupled and in best cases all variables can be expressed with analytic formulas.}  
\item{The linear algebraic equation system for the exponents are under-determined, 
leaving usually one self-similar exponent completely free, which means an extra free parameter in the 
obtained ODE system, causing a very rich mathematical structure. The free exponent can have either positive or negative sign. Negative values usually result in power-law divergent or exploding solutions in contrary, positive exponents mean power-law decaying solutions which are desirable for dissipative systems.}
\end{itemize}

In our former works we mostly analyzed viscous fluids like incompressible or compressible viscous Newtonian fluids where all exponents had positive values, which is also true for the regular diffusion process, where $\alpha$ and $\beta = 1/2$. In our recent case these values are rather expected to be negative %signs 
since the observed, approximately-flat Eucledian Universe scenario.
 
Our present ODE system of equations~\eqref{ODE} cannot be separated and solved with analytic means. Non-autonomous ODE systems can be linearized and the stationer points can be found which helps to sketch the global qualitative properties of the solutions in the phase space. No such mathematical method exists for non-autonomous non-linear systems.  Therefore there is no general tool in our hands that could help us to determine the global properties of the fluid. 
The equations of the stationary points however can be given as
$f'(\omega) = 0$ and $g'(\omega) = 0 $, then we obtain
\begin{eqnarray}
g\omega + 2fg = 0, \nonumber \\ 
g^2 = 0. 
\end{eqnarray} 
From the first equation for $g \ne 0$ case we get that the $ f = -\frac{\omega}{2} $ function contains the stationary points, with the dimensional definition of velocity, $r/2t$, with lack of information on the general properties of the solutions. From the second equation one can get that $g = 0$ is stationary point, which is a trivial statement meaning that zero density is a stationery point. 

We do not even know if this ODE system has mathematically rigorous existence and unicity theorem for the solutions. 
The only physically reasonable way that we can follow is to performed numerous calculations with large number of parameter sets to explore the properties of the solutions in wide range.  
Our numerical experiences say that both field variable has continuous solution on a closed 
interval on the half axis of time and radial distance. We found no compact supports or 
ruptures in the solutions. 

%%%%%%%%%%%%%%%%%%%%%%%%%%%%%%%%%%%%%%%%%%%%%%%%%%%%%%%%%%%%%%%%%%%%%%
\section{Results}
\label{sec:results}

To understand the physics of the presented model a systematic parameter study of eq.~\eqref{ODE} is required. 
First we investigate the role of the $w$ which is the strength of the linear equation of state as well.  
 
\begin{figure}  
\begin{center}
\scalebox{0.35}{
\rotatebox{0}{\includegraphics{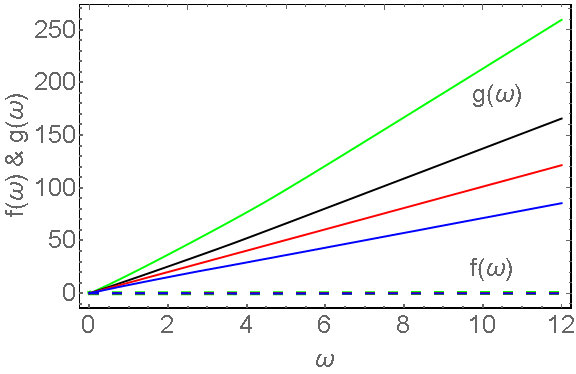}}}
\end{center}
\caption{The shape functions of Eq.~\eqref{ODE} for different EoS strength parameters: $ w = -1$ (green), $-0.33$ (black), $0$ (red), and $0.33$ (blue). The solid and dashed lines represent shape functions of density $g(\omega)$ and velocity $f(\omega)$, respectively. Note, all velocity curves coincide.
The initial conditions are $f(10^{-3}) = 0.5$ and $g(10^{-3}) = 0.01$.}	
\label{kettes}        
\end{figure}

Figure~\ref{kettes} presents the density and velocity shape functions for four physically relevant different EoS strength parameters, $ w = -1$, $-0.33$, $0$, and $0.33$. We integrate the ODE system between $ \omega_{min} = 1 \cdot 10^{-3}$ 
and $ \omega_{max} = 5 \cdot 10^{1}$.  
As initial conditions for velocity and density we took  $f(\omega_{min}) = 0.5$ and $ g(\omega_{min}) = 0.01$.  The ratio $f(\omega_{min})/g(\omega_{min})$ was set to 50 here. This choice of the initial velocity (a positive value) means an initially radially expanding fluid, and a physically reasonable positive density. We found, that all four {\sl dashed} $f(\omega)$ velocity curves are constant and indistinguishable at this range and line widths. All four {\sl solid}, $g(\omega)$ density curves have a linear dependence of $\omega$. 

From now on we fix $w = -1$ which is the choice of the dark matter.  
%
%present not much change for $w \leq 0$ values and all %remain more-or-less constant. This is also true for the %$w> 0$ values up to $\omega \approx 10$, however beyond %it result in linear dependence of $\omega$. The density %shape functions are all look constants, $g(\omega)=0.1$ %for all the investigated $w$ values. The presented four %{\sl solid} curves are indistinguishable at this range %and line widths. 
%
%%%%%%%%%%%%%%%%%%%%%%%%%%%%%%%%%%%%%%%%%%%%%%%%%%%%
%%%%%%%%%%%%%%%%%%%%%%%%%%%%%%%%%%%%%%%%%%%%%%%%%%%%
%
%To understand this result, one can see the right hand %side of the Euler equation in eq.~\eqref{ODE}. Here the %first term is inherited by the EoS, and it is %proportional to $g'/g$ which is much smaller for small %$\omega$ arguments than the last two %gravitation-related terms. In other words the %gravitation term is much more relevant than the %pressure term, which also means high-level of stability %of our model against the perturbation of the EoS. 
%
%%%%%%%%%%%%%%%%%%%%%%%%%%%%%%%%%%%%%%%%%%%%%%%%%%%%
\begin{figure*}
\begin{center}
%\scalebox{0.25}{\rotatebox{0}{\includegraphics{density_3d.png}}}
%\scalebox{0.25}{\rotatebox{0}{\includegraphics{velocity_3d.png}}}
\scalebox{0.3}{\rotatebox{0}{\includegraphics{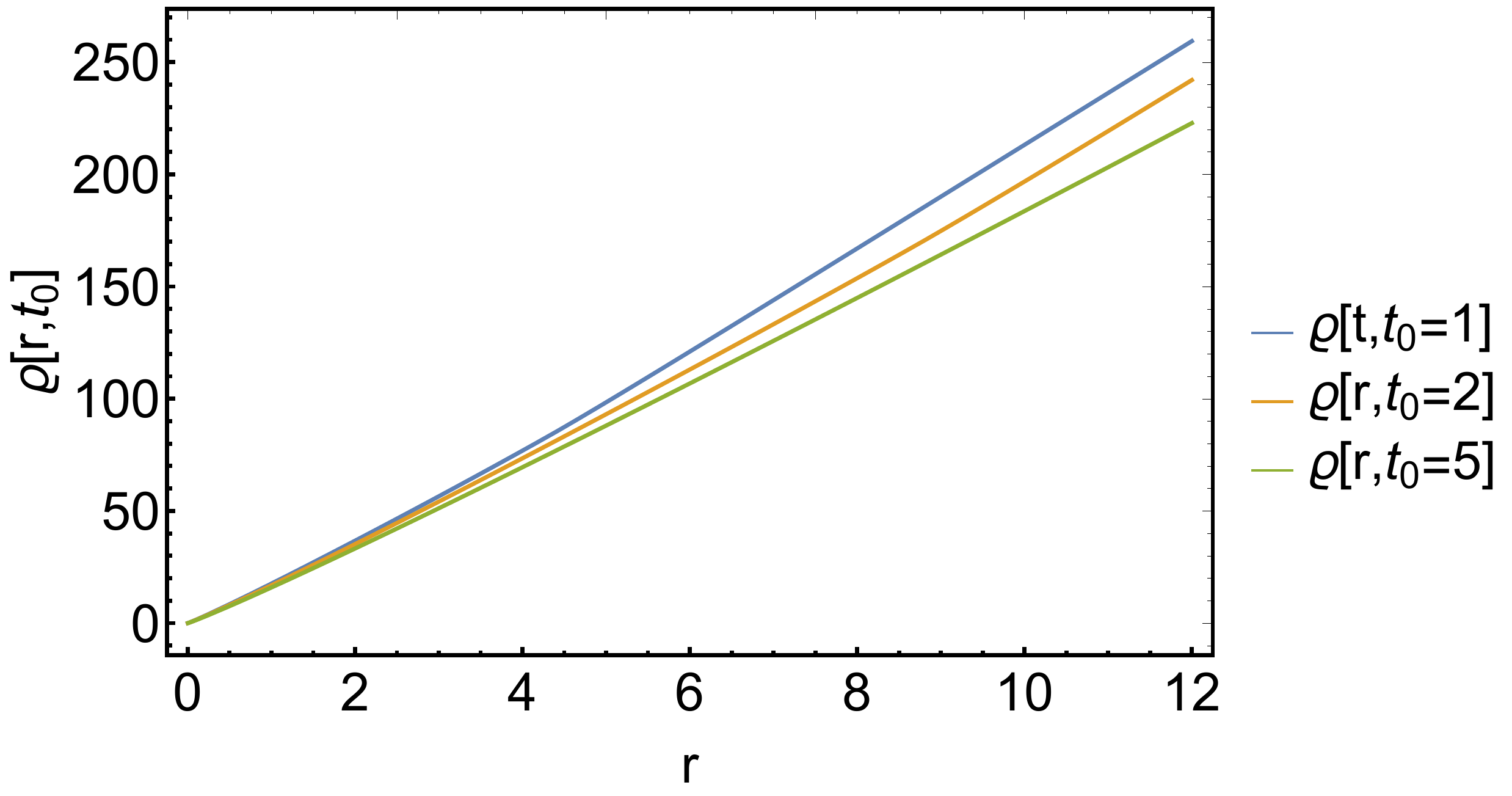}}}
\scalebox{0.305}{\rotatebox{0}{\includegraphics{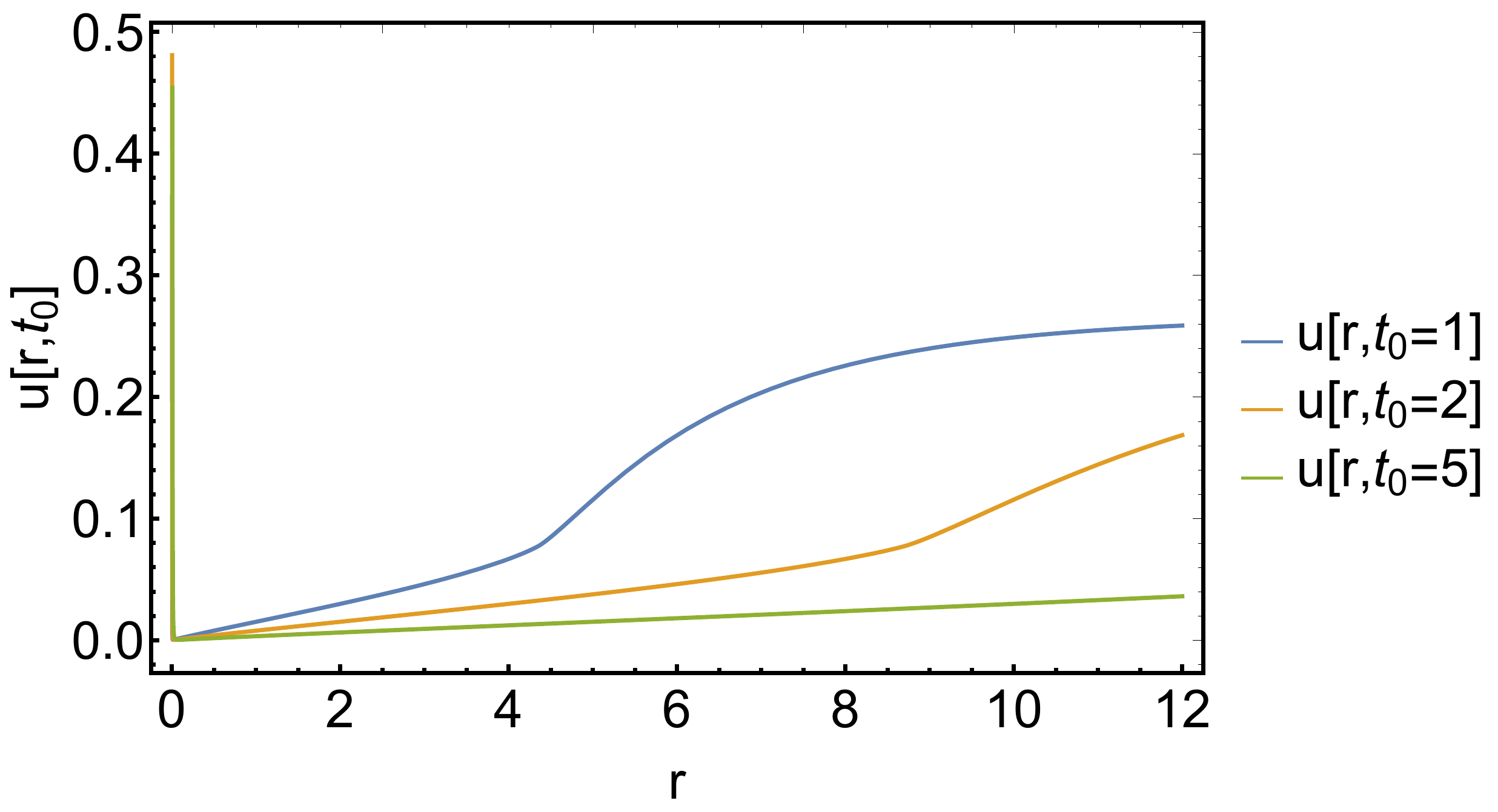}}}
\scalebox{0.3}{\rotatebox{0}{\includegraphics{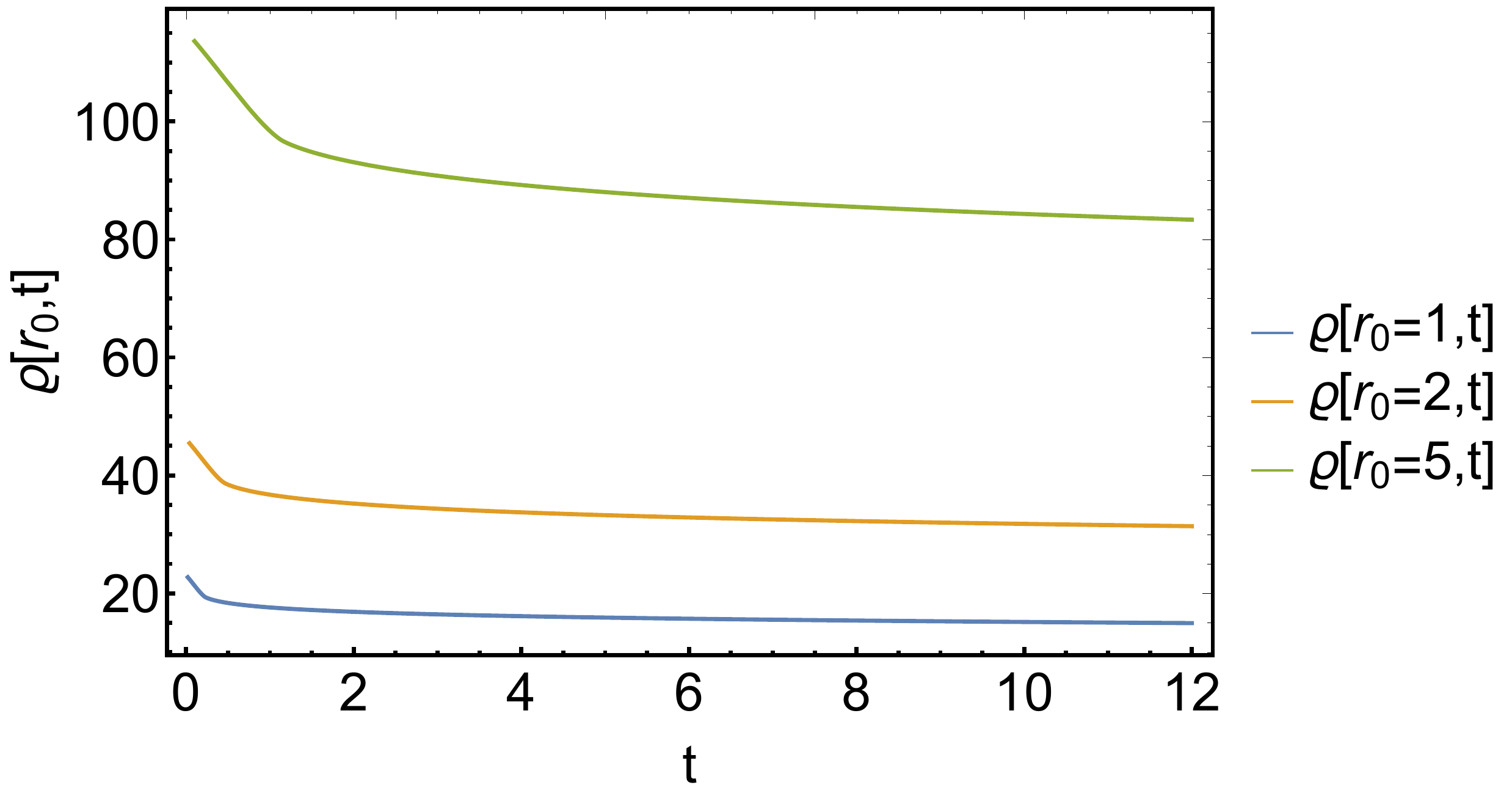}}}
\scalebox{0.305}{\rotatebox{0}{\includegraphics{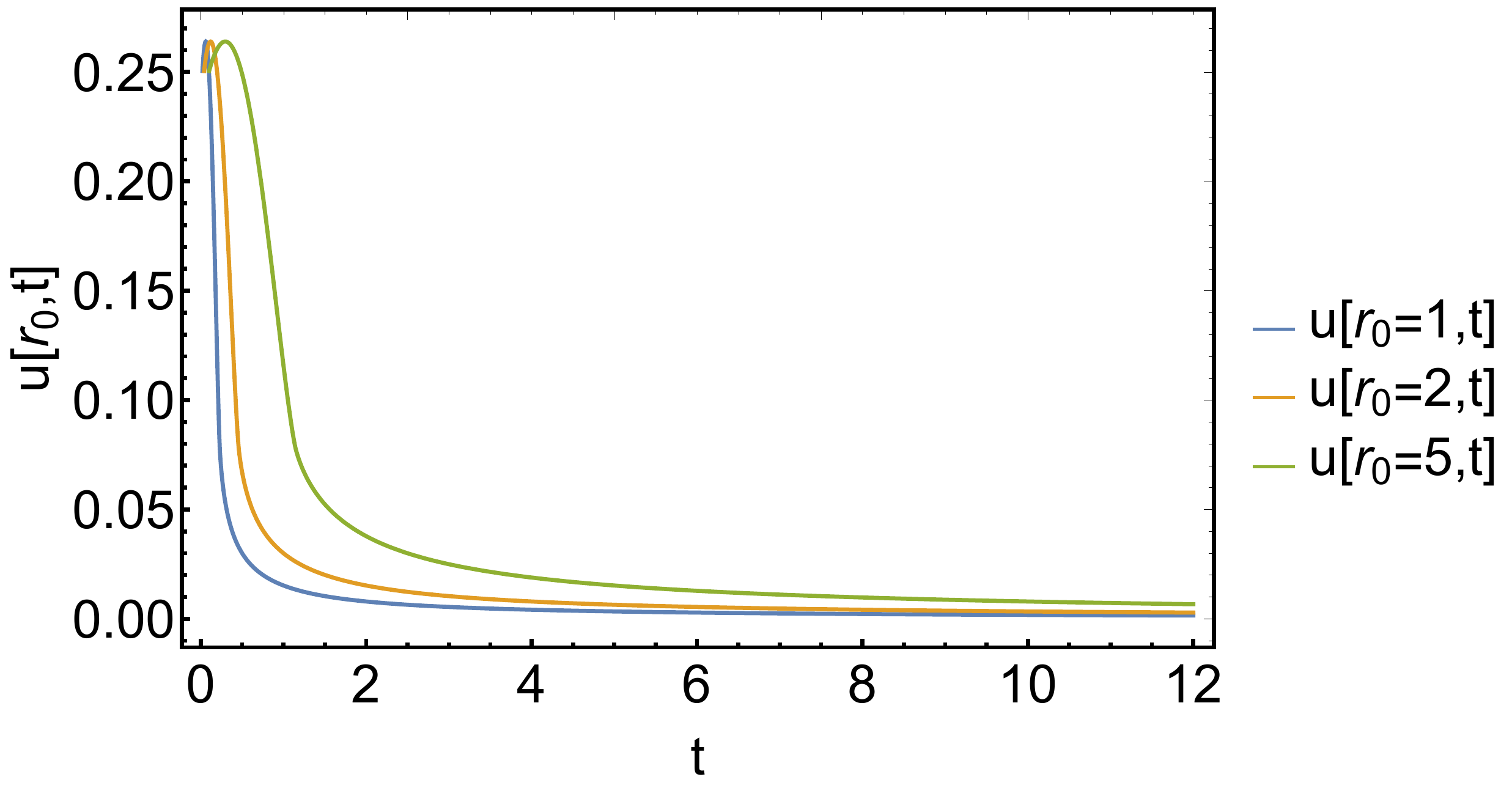}}}
\caption{The space-time evolution of the density (left) and the velocity (right) distributions %functions 
for the case, $w = -1$. Curves are for the radial (top panels) and time (bottom panels) dependence at different fixed time- and space points. %, respectively.  
The initial conditions are $f(10^{-3}) = 0.5$ and $g(10^{-3}) = 0.01$.}
\label{harmas}     
\end{center}
\end{figure*}
%%%%%%%%%%%%%%%%%%%%%%%%%%%%%%%%%%%%%%%%%%%%%%%%%%% 
%
Figure~\ref{harmas} shows the space and time projections of the obtained density and the radial velocity distributions in natural units. {\sl Upper panels} are the radial (space) evolution of the density ({\sl left}) and radial velocity ({\sl right}) functions. {\sl Lower panels} present the time-projections of the above distributions at fixed radius values, respetively.

The density has linear dependence of the radius at any time. The radial dependencde of velocity function looks however a bit more different: at very small distances close to the origin for all time points it has a sharp down running edge with a minima and a linear dependece at larger distances. 

The time dependeces are different. The density has a quick but linear decay 
at small times and another but more slower (and almost linear) decay at larger 
times. The velocity shape functions start at non-zero values and have a quick 
deacy in time at all distances. These properties cannot be anticipated directly  from the shape functions. We note, these are the direct results of our presented model with initial conditions for the same point, $f(\omega_{min}=10^{-3}) = 0.5$ and $g(\omega_{min}=10^{-3}) = 0.01$. 
 
It is more relevant to investigate the dynamics of the complete fluid in time and space to understand some general trends or physical phenomena as the function of the initial conditions. For this reason one can calculate the total energy density of the system including the kinetic and potential terms, 
\begin{equation}
E_{tot} = E_{kin} + E_{pot} = \frac{1}{2}u^2 \rho - \frac{\rho}{r} \ . 
\end{equation}

\begin{figure}
\begin{center}
\scalebox{0.25}{\rotatebox{0}{\includegraphics{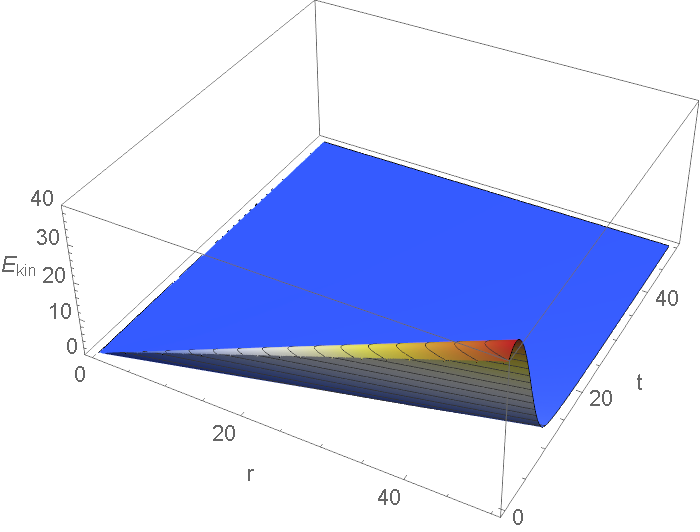}}}
\scalebox{0.25}{\rotatebox{0}{\includegraphics{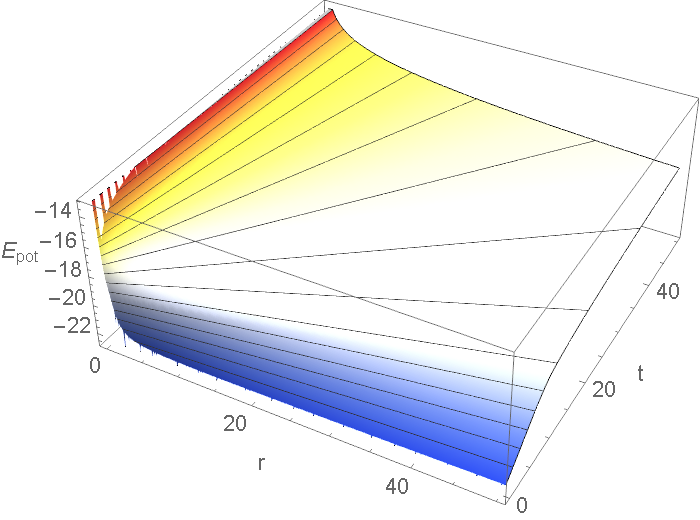}}}
%\scalebox{0.27}{\rotatebox{0}{\includegraphics{Etot_3d.pdf}}}
\end{center}
\caption{The kinetic and potential energy densities of the investigated fluid for $w = -1 $, with  
initial conditions $f(10^{-3}) = 0.5$ and $g(10^{-3}) = 0.01$.} 	
\label{hatos}       
\end{figure}

Figure~\ref{hatos} presents the kinetic, $E _{kin}$ and the potential energy, $E_{pot}$ densities -- the two reasonable physical quantities which can be evaluated from the hydrodynamical model for a given parameter set. 
The kinetic energy is zero in the origin, has a linearly enhancing maxima 
at larger distances and has a quick decay in time for all radial distances. This means parctically a blowing up sphere where the fluid moves. 
The potential energy has different properties, finitie values in the origin at 
practically all times, and a slow decay for large times and distances. 

% Both functions have the same general feature, a quick decay for large %distances at any time point within the same domain. For the given initial %conditions the kinetic energy density is a factor of five larger, than the %potential energy densities. Therefore the total energy density is not presented, %since it inherited the same features from the dominating kinetic energy density. 
%%%%%%%%%%%%%%%%%%%%%%%%%%%%%%%%%%%%%%%%%%%%%%%%%%%%

Since the solutions of our non-autonomous non-linear ODE system is not known, we investigated how the solutions are depend on the initial conditions. By our analysis different physical scenarios available if we drastically change the ratio of the initial velocity and density. Another point is the effect of the changing the sign of the initial velocity. Positive sign means expansion, which may be stopped later, while negative sign result in an initial collapse of the fluid.  

%%%%%%%%%%%%%%%%%%%%%%%%%%%%%%%%%%%
\begin{figure*}[!h]
\begin{center}
%\scalebox{0.25}{\rotatebox{0}{\includegraphics{k_mn.png}}}
%\scalebox{0.25}{\rotatebox{0}{\includegraphics{tot_mn.png}}}
\scalebox{0.25}{\rotatebox{0}{\includegraphics{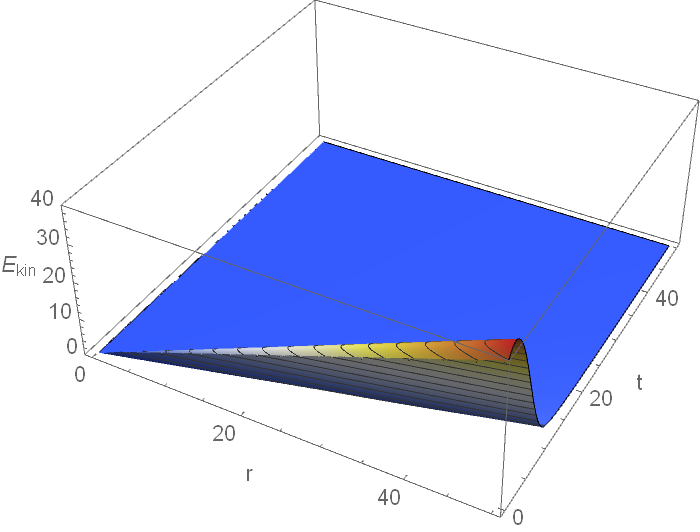}}}
\scalebox{0.25}{\rotatebox{0}{\includegraphics{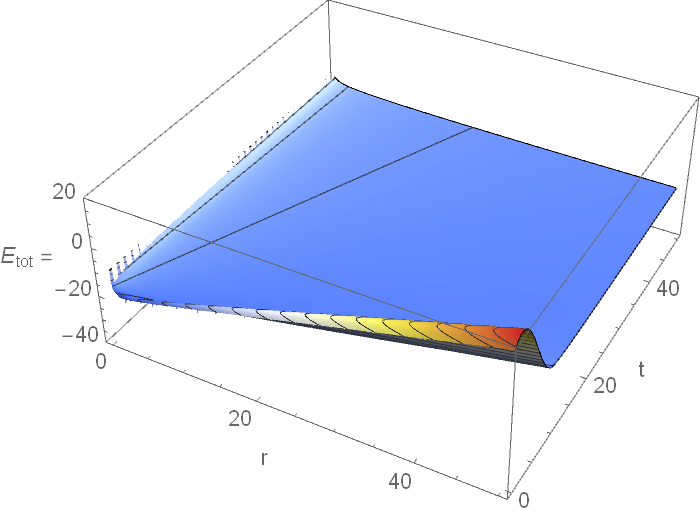}}}
\scalebox{0.25}{\rotatebox{0}{\includegraphics{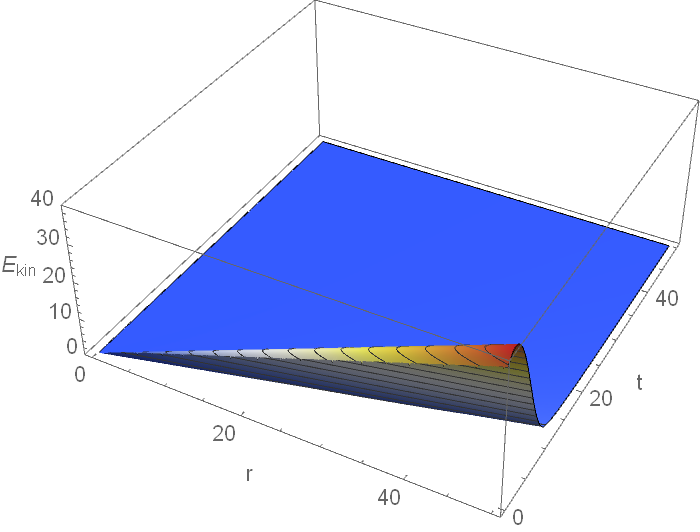}}}
\scalebox{0.25}{\rotatebox{0}{\includegraphics{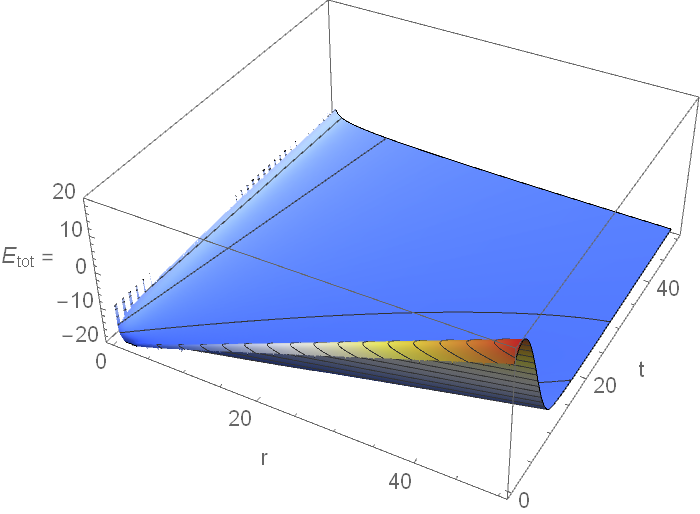}}} 
\scalebox{0.25}{\rotatebox{0}{\includegraphics{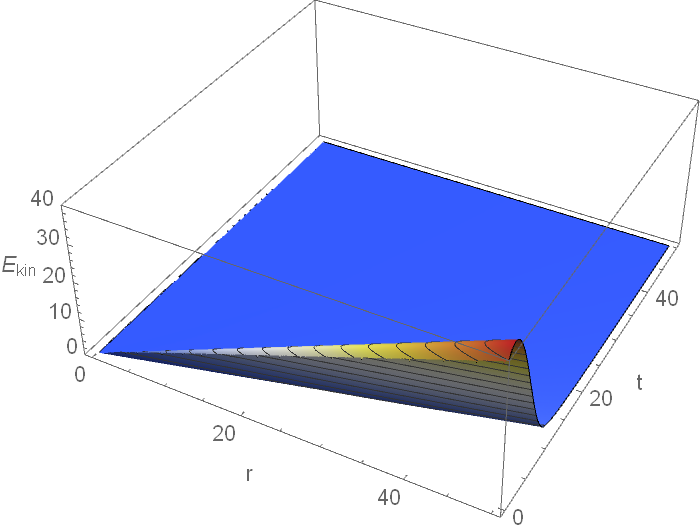}}}
\scalebox{0.25}{\rotatebox{0}{\includegraphics{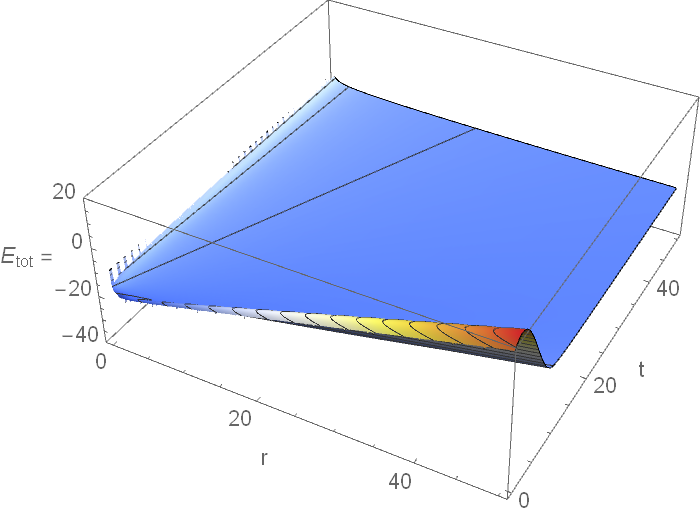}}}
%\scalebox{0.25}{\rotatebox{0}{\includegraphics{k_pn.png}}}
%\scalebox{0.25}{\rotatebox{0}{\includegraphics{tot_pn.png}}}
\caption{The first column presents the kinetic energies and the second is for the total energies calculated with different initial velocities for the same initial density of $g(10^{-3}) = 10^{-1}$, and for $ w = -1$. 
The five following rows show the corresponding physical quantities for $f(10^{-3}) =-0.5$, $0$, and $0.5$ initial velocities, respectively.} 
\label{hetes}     
\end{center}
\end{figure*}
%%%%%%%%%%%%%%%%%%%%%%%%%%%%%%%
%\newpage
Figure~\ref{hetes} depicts three %five
different kinetic and total energy density distributions of the fluid 
under each other for different $  f(\omega_{min}/g(\omega_{min}) $ ratios. 
The initial density is fixed to 0.01 for all five cases.  
For a better comparison the both the time and the space domains were taken: $[0,50$] in natural units for all panels. The chosen ranges are varies, because it is not possible to present all functions in the same range in an informative manner.  
   
\begin{description}
%\item[$ f(\omega_{min}) = -15 $] results presented in the {\sl first line panels}, where the numerical ration of the two initial conditions is $-1500$. This scenario can be defined as collapse which is started with a  the initial kinetic energy density that suppresses the potential energy density. 
%The general features are the following, the origin or center of the fluid gain and more kinetic and total energy densities as time goes by. This can be explained with $\gamma = -1$ negative self-similar exponent, which means an exploding density solution. At small times the major part of the fluid has no total energy density at all. As time goes by the kinetic and total energy densities of the fluid slowly expands. The shape of the expansion front become more and more diffuse at larger time points thanks to the $\beta = 1$ self-similar exponent which is responsible for dispersion. 
%The case with large positive initial velocity value, $ f(\omega_{min}) = +15 $ result the same function in the {\sl bottom line panels}, due to the squared velocity in the kinetic term. Thus it is completely indifferent if the fluid is strongly pushed together or strongly pulled apart -- it will gain infinite kinetic energy as time goes to infinity. This is a highly unusual feature of our model which contradicts our physical intuition.

\item[$ f(\omega_{min}) = -0.5$] case is seen in the {\sl first %second
line panels} with setting the kinetic-to-potential energy ratio to -50. Both physical quantity changed radically. The feature that the center of the fluid gains more and more kinetic and potential energy density is much more suppressed, on the other side at small times the kinetic energy density of the total fluid enhances. The larger the radius the larger the energy density gain. Note, that the values of the maximum linearly enhance with time. 
We may say, that this phenomena is a kind of "delayed" acceleration and deceleration of the fluid due to the enhanced potential energy density. This effect is not visible for larger initial velocities, where the initial kinetic energy densities dominates the potential energy. In this scenario the strength of the two energy density terms lies in almost in the same range. The figure of the total energy density function is drastically different to the former case, since it become negative everywhere. The effect of the kinetic energy density for small time values at any radius are clearly seen as a "bump" which lifts the potential energy density. 
The same statements are valid also for the panel of the {\sl third %fourth
line panels},  where the dynamics for small positive initial velocity case, $ f(\omega_{min}) = +0.5 $  is presented.

\item[$ f(\omega_{min}) = 0 $] case describes the dynamics of a fluid with zero initial velocity ({\sl middle panels}). Here the effect of the potential energy density can be studied in a non-perturbed way. The "bump" of the kinetic energy color density for small times at all fluid radius -- a short acceleration and deceleration -- of the fluid is originated by the gravitational potential of the fluid itself. 

\end{description}
We must summarize our results as follows: for this ideal hydrodynamical model, which does not include any kind of viscous damping, the self-similar {\sl ansatz} may give solutions which explode in time, if the initial conditions are so.  

%%%%%%%%%%%%%%%%%%%%%%%%%%%%%%%%%%%%%%%%%%%%%%%%%%%%%%%%%%%%%%%%%%%%%%
\section{Discussion}
\label{sec:dis}

The analytical solution was motivated by the cosmological observations.
The Hubble-law provides a velocity scaling~\cite{HubbleLaw} for the expansion of the Universe, numerous possible scaling mechanisms are coded in our model via the choice of the self-similar {\sl ansatz} several exponents can describe various 
time decay scenarios.  As Figure~\ref{kettes} presented a linear scale parameter is obtained.
%below and above a critical scaling value, $\omega \approx 10$. At this critical point the quick change in the scaling is provided naturally as an in-coded inflation-like behavior. 
Although, the scaling curves are linear everywhere, at around a certain $r$ values a quick change in the velocity is provided naturally as an in-coded inflation-like behavior by the {\sl ansatz} . This critical $r$ value depends on the initial condition.
As {\sl right panels} of Figure~\ref{harmas} shows high initial velocity in the dark-fluid, $w=-1$ limit (iii), will relax to a small, constant 'non-relativistic' value at long timescale. Meanwhile, the dependence on the distance is getting saturated and close to flat at far distances from the initial 'Big Bang' point. We have found, that it is $\lesssim 1\%$ effect for nearby objects. This precision however, well beyond the uncertainty of the measured Hubble-constant value~\cite{Hubble,  WAMP09, Planck, Ligo} 
All $u[r_0, t]$ curves in the lower right part of Figure~\ref{harmas} are getting fully flat beyond $r = 2-4$. 

The solution for the local density scaling is increasing in space but flat in time. This results the space-time density evolution function in the {\sl left panels} of Figure~\ref{harmas}. Around the center  this function starts linearly with the distance, and after a rapid rise, the density of the fluid is getting flat.  %However an observer located around $r=2-4$ units, would also see homogeneous matter within a few percent. 
An interesting feature of the model, that apart from the initial point, the local density is increasing with time -- due to the dispersing shock-wave and the dark matter equation of state ($w=1$).

The structure of the ODE and the choice of the initial conditions leave us enough freedom for the valid physical picture. As we could see, the $E_{tot}\approx 0$ is well supported by the solution, especially in the infinite space-time limit presented in Figure~\ref{hetes}.

One may also set the initial condition according to that the Universe is observed as flat Euclidean, with $\Omega=\overline{\rho}_i/\rho_c \approx 1$, and indeed,  
\begin{equation}
\Omega = \Omega_{M}+ \Omega_{\Lambda} \ \ \textrm{where} \ \ \Omega_{M}=\Omega_{B}+ \Omega_{CDM} \ . 
\end{equation}
Here $\Omega_i=\overline{\rho}_i/\rho_c$ for '$i$': baryonic ($B$) and cold-dark matter ($CDM$), as well as dark energy ($\Lambda$), respectively. Thus, we can set the asymptotic flat  value for the ratio of the kinetic and total energy of the Universe~\cite{Dimitar,Peacock,Hinshaw}, 
\begin{equation}
E_{kin}/E_{tot} \sim \Omega_{M}/\Omega \approx 0.27 \ . 
\end{equation} 
Seemingly, our solution can be compatible with a realistic cosmological picture with fine-tuned initial conditions, which can draw further consequences of the evolution of a early dark-fluid Universe.
For $f(\omega_0 =0.01) = 0.1 $ and $g(\omega_0 = 0.01) = 0.1 $ initial condition we get an $ E_{kin}/E_{tot} = 0.26 $ ratio 
for $r = 58 $ and $t = 2$ spatial coordinate and time point, which 
can be scaled to the present status of the Universe.
 
%%%%%%%%%%%%%%%%%%%%%%%%%%%%%%%%%%%%%%%%%%%%%%%%%%%%%%%%%%%%%%%%%%%%%%%%%%%%%%%%%%
\section{Summary and outlook} 
\label{sec:sum}
We presented and analyzed a model where the spherically symmetric compressible Euler equation 
was used with a general linear matter equation of state and gravitation. This model with $w=-1$ is the simplest candidate for dark fluid. As a method of investigation we applied the self-similar {\it ansatz},  which might fullfills the expected ratio of $E_{kin}/E_{tot} \approx 0.27$ depending on the choice of initial conditions. %Certainly, this model can be improved according to the recent standard cosmology picture.
 
Further work is in progress to improve our hydrodynamical model, 
e.g. including fluctuations~\cite{CsabaiP}, rotation, relativistic or two-fluid effects~\cite{twofluid1,twofluid2} or quantum mechanical effects~\cite{madelung}  in the near future.  
%%%%%%%%%%%%%%%%%%%%%%%%%%%%%%%%%%%%%%%%%%%%%%%%%%%%%%%%%%%%%%%%%%%%%%%%
\section*{Acknowledgments}

This work was supported by Hungarian National Research Fund (OTKA) grants K123815, K135515
NKFIH 2019-2.1.11-TÉT-2019-00050, 2019-2.1.11-TÉT-2019-00078, THOR COST action CA15213, and the Wigner GPU Laboratory.
 
%\end{acknowledgments}
%%%%%%%%%%%%%%%%%%%%%%%%%%%%%%%%%%%%%%%%%%%%%%%%%%%%%%%%%%%%%%%%%%%%%%%%%%%%%%%%%%
{}
\end{document}